\documentclass{elsart}
\usepackage{ifpdf}
\usepackage{graphicx,amssymb,lineno,epsfig,subfig}
\usepackage{subfig}
\usepackage{indentfirst}
\ifpdf
\usepackage[%
  pdftitle={Instructions for use of the document class
    elsart},%
  pdfauthor={Simon Pepping},%
  pdfsubject={The preprint document class elsart},%
  pdfkeywords={instructions for use, elsart, document class},%
  pdfstartview=FitH,%
  bookmarks=true,%
  bookmarksopen=true,%
  breaklinks=true,%
  colorlinks=true,%
  linkcolor=blue,anchorcolor=blue,%
  citecolor=blue,filecolor=blue,%
  menucolor=blue,pagecolor=blue,%
  urlcolor=blue]{hyperref}
\else
\usepackage[%
  breaklinks=true,%
  colorlinks=true,%
  linkcolor=blue,anchorcolor=blue,%
  citecolor=blue,filecolor=blue,%
  menucolor=blue,pagecolor=blue,%
  urlcolor=blue]{hyperref}
\fi

\renewcommand{\theequation}{\arabic{section}.\arabic{equation}}

\def\@subjclass{}
\makeatletter
\def\elsartstyle{%
    \def\normalsize{\@setfontsize\normalsize\@xiipt{14.5}}
    \def\small{\@setfontsize\small\@xipt{13.6}}
    \let\footnotesize=\small
    \def\large{\@setfontsize\large\@xivpt{18}}
    \def\Large{\@setfontsize\Large\@xviipt{22}}
    \skip\@mpfootins = 18\p@ \@plus 2\p@
    \normalsize
}
\@ifundefined{square}{}{\let\Box\square}
\makeatother
\usepackage{indentfirst}
\usepackage{setspace}
\begin{document}

\begin{frontmatter}
\title{A type of bounded traveling wave solutions for the Fornberg-Whitham equation}
\author{Jiangbo Zhou\corauthref{cor}},
\corauth[cor]{Corresponding author.} \ead{zhoujiangbo@yahoo.cn}
\author{Lixin Tian}
\address{Nonlinear Scientific Research Center, Faculty of Science, Jiangsu University,
Zhenjiang, Jiangsu, 212013, China}
\begin{abstract}
In this paper, by using bifurcation method, we successfully find the
Fornberg-Whitham equation
\[
u_t - u_{xxt} + u_x = uu_{xxx} - uu_x + 3u_x u_{xx},
\]
\noindent has a type of traveling wave solutions called kink-like
wave solutions and antikink-like wave solutions. They are defined on
some semifinal bounded domains and possess properties of kink waves
and anti-kink waves. Their implicit expressions are obtained. For
some concrete data, the graphs of the implicit functions are
displayed, and the numerical simulation is made. The results show
that our theoretical analysis agrees with the numerical simulation.
\end{abstract}


\end{frontmatter}
\section{Introduction}
 \setcounter {equation}{0}
It is well known that the exact solutions for the nonlinear partial
differential equations can help people know deeply the described
process. So an important issue of the nonlinear partial differential
equations is to find their new exact solutions. Traveling wave
solution is an important type of solution for the nonlinear partial
differential equations and many nonlinear partial differential
equations have been found to have a variety of traveling wave
solutions. For instances, the well-known Korteweg-de Vries equation

\begin{equation}
\label{eq1} u_t - 6uu_x + uu_{xxx} = 0
\end{equation}
\noindent has solitary wave solutions and its solitary waves are
solitons \cite {1}. A KdV-like equation
\begin{equation}
\label{eq2} u_t + a(1 + bu^n)u^nu_x + \delta u_{xxx} = 0
\end{equation}
\noindent has some kink wave solutions \cite {2}. Its kink wave
solution $u(\xi )(\xi = x - ct)$ was difined on $( - \infty , +
\infty )$, and $\mathop {\lim }\limits_{\xi \to - \infty } u(\xi ) =
A$, $\mathop {\lim }\limits_{\xi \to \infty } u(\xi ) = B$, where
$A$ and $B$ are two constants and $A \ne B$. The Camassa-Holm
equation
\begin{equation}
\label{eq3} u_t - u_{xxt} + 3uu_x = 2u_x u_{xx} + uu_{xxx}
\end{equation}
\noindent has peakons, cuspons, stumpons, composite wave solutions
\cite {3}, it also has \noindent compactons \cite {4}. The
Degasperis-Procesi equation
\begin{equation}
\label{eq4} u_t - u_{xxt} + 4uu_x = 3u_x u_{xx} + uu_{xxx}
\end{equation}
\noindent has a multitude of peculiar wave solutions: peakons,
cuspons, composite waves, and stumpons \cite {5}. The
Kuramoto-Sivashinsky equation
\begin{equation}
\label{eq5} u_t + uu_x + u_{xx} + u_{xxxx} = 0
\end{equation}
\noindent has periodic and solitary solutions \cite {6}. In \cite
{7}, Liu, Li and Lin found a new type of traveling wave solutions
for the Camassa-Holm equation, which are defined on some semifinal
bounded domains and possess properties of kink waves or anti-kink
waves. They called them kink-like waves and antikink-like waves.
Later, Guo and Liu \cite {8}  found the CH$ - \gamma $ equation
\begin{equation}
\label{eq6} u_t + c_0 u_x + 3uu_x - \alpha ^2(u_{xxt} + uu_{xxx} +
3u_x u_{xx} ) + \gamma u_{xxx} = 0,
\end{equation}
\noindent posses kink-like waves when $\alpha ^2>0$. Tang and Zhang
\cite{9} showed the CH$ - \gamma $ equation (\ref{eq6}) also has
kink-like wave solutions even when $\alpha ^2<0$. Chen and Tang
\cite {10} showed that the Degasperis-Procesi equation (\ref{eq4})
has such type of traveling wave solutions. Recently, Liu and Yao
\cite{11} found the following generalized Camassa-Holm equation
\begin{equation}
\label{eq7} u_t + 2ku_x - u_{xxt} + auu_x = 2u_x u_{xx} + uu_{xxx}
\end{equation}
also posses kink-like wave solutions.

We are motivated to seek kink-like wave and antikink-like wave
solutions for the Fornberg-Whitham equation
\begin{equation}
\label{eq8} u_t - u_{xxt} + u_x = uu_{xxx} - uu_x + 3u_x u_{xx} .
\end{equation}
To our knowledge, such type of traveling wave solution has never
been found for the Fornberg-Whitham equation. Eq.(\ref{eq8}) was
used to study the qualitative behaviours of wave-breaking \cite
{12}. It admits a wave of greatest height, as a peaked limiting form
of the traveling wave solution \cite {13}, $u(x,t) = A\exp
(-\frac{1}{2}\left| {x - \frac{4}{3}t} \right|)$, where $A$ is an
arbitrary constant.

The remainder of the paper is organized as follows. In Section 2, we
state the main results which are implicit expressions of the
kink-like wave and the antikink-like wave solutions. In Section 3,
we give the proof of the main results. In Section 4, we make the
numerical simulation of the kink-like and the antikink-like waves. A
short conclusion is given in Section 5.

\section{Main results}

 \setcounter {equation}{0}
 We state our main result as follows.

\noindent \textbf{Theorem 1. }For given constant $c$, let
\begin{equation}
\label{eq9} \xi = x - ct
\end{equation}
\begin{equation}
\label{eq10} \varphi _0^\pm = c - 1\pm \sqrt {(c - 1)^2 - 2g}
\end{equation}
\begin{equation}
\label{eq11} g_1 (c) = \frac{(c - 1)^2}{2}
\end{equation}
\begin{equation}
\label{eq12} g_2 (c) = \frac{(c - 1)^2 - 1}{2}
\end{equation}
\noindent (1) If $g < g_2 (c)$, then Eq.(\ref{eq8}) has two
kink-like wave solutions $u = \varphi _1 (\xi )$ and $u = \varphi _3
(\xi )$ and two antikink-like wave solutions $u = \varphi _2 (\xi )$
and $u = \varphi _4 (\xi )$.
\begin{equation}
\label{eq13} \frac{(2\sqrt {\varphi _1^2 + l_1 \varphi _1 + l_2 } +
2\varphi _1 + l_1 )(\varphi _1 - \varphi _0^ - )^{\alpha _1
}}{(2\sqrt {a_1 } \sqrt {\varphi _1^2 + l_1 \varphi _1 + l_2 } + b_1
\varphi _1 + l_3 )^{\alpha _1 }} = \beta _1 e^{ - \frac{1}{2}\xi
},\\
\quad \xi \in ( - \infty ,\xi _0^1 )
\end{equation}
\begin{equation}
\label{eq14} \frac{(2\sqrt {\varphi _2^2 + l_1 \varphi _2 + l_2 } +
2\varphi _2 + l_1 )(\varphi _2 - \varphi _0^ - )^{\alpha _1
}}{(2\sqrt {a_1 } \sqrt {\varphi _2^2 + l_1 \varphi _2 + l_2 } + b_1
\varphi _2 + l_3 )^{\alpha _1 }} = \beta _1 e^{\frac{1}{2}\xi },\\
\quad \xi \in ( - \xi _0^1 ,\infty )
\end{equation}
\begin{eqnarray}
\label{eq15}  \lefteqn{  \frac{(2\sqrt {\varphi _3^2 + m_1 \varphi
_3 + m_2 } + 2\varphi _3 + m_1 )(\varphi _3 - \varphi _0^ +
)^{\alpha _2 }}{(2\sqrt {a_2 } \sqrt {\varphi _3^2 + m_1 \varphi _3
+ m_2 } + b_2 \varphi _2 + m_3 )^{\alpha _2 }} = \beta _2 e^{ -
\frac{1}{2}\xi }, \xi \in ( - \xi _0^3 ,\infty )}
\nonumber\\
& &
\end{eqnarray}
\noindent and
\begin{eqnarray}
\label{eq16}  \lefteqn{ \frac{(2\sqrt {\varphi _4^2 + m_1 \varphi _4
+ m_2 } + 2\varphi _4 + m_1 )(\varphi _4 - \varphi _0^ + )^{\alpha
_2 }}{(2\sqrt {a_2 } \sqrt {\varphi _4^2 + m_1 \varphi _4 + m_2 } +
b_2 \varphi _4 + m_3 )^{\alpha _2 }} = \beta _2 e^{\frac{1}{2}\xi },
\xi \in ( - \infty ,\xi _0^3 )}
\nonumber\\
& &
\end{eqnarray}
\noindent where
\begin{equation}
\label{eq17} l_1 = \frac{2}{3}(1 - 3c - 3\sqrt {(c - 1)^2 - 2g} )
\end{equation}
\begin{equation}
\label{eq18} l_2 = \frac{2}{3}[1 - 4c + 3c^2 - 3g + (3c + 1)\sqrt
{(c - 1)^2 - 2g} ]
\end{equation}
\begin{equation}
\label{eq19} l_3 = \frac{4}{3}[2 - 5c + 3c^2 - 6g + (3c + 2)\sqrt
{(c - 1)^2 - 2g} ]
\end{equation}
\begin{equation}
\label{eq20} m_1 = \frac{2}{3}(1 - 3c + 3\sqrt {(c - 1)^2 - 2g} )
\end{equation}
\begin{equation}
\label{eq21} m_2 = \frac{2}{3}[1 - 4c + 3c^2 - 3g - (3c + 1)\sqrt
{(c - 1)^2 - 2g} ]
\end{equation}
\begin{equation}
\label{eq22} m_3 = \frac{4}{3}[2 - 5c + 3c^2 - 6g - (3c + 2)\sqrt
{(c - 1)^2 - 2g} ]
\end{equation}
\begin{equation}
\label{eq23} a_1 = 4(1 - 2c + c^2 - 2g + \sqrt {(c - 1)^2 - 2g} )
\end{equation}
\begin{equation}
\label{eq24} a_2 = 4(1 - 2c + c^2 - 2g - \sqrt {(c - 1)^2 - 2g} )
\end{equation}
\begin{equation}
\label{eq25} b_1 = - \frac{4}{3} - 4\sqrt {(c - 1)^2 - 2g}
\end{equation}
\begin{equation}
\label{eq26} b_2 = - \frac{4}{3} + 4\sqrt {(c - 1)^2 - 2g}
\end{equation}
\begin{equation}
\label{eq27} \alpha _1 = - \frac{1 + \sqrt {(c - 1)^2 - 2g} }{2\sqrt
{(c - 1)^2 - 2g + \sqrt {(c - 1)^2 - 2g} } }
\end{equation}
\begin{equation}
\label{eq28} \alpha _2 = \frac{ - 1 + \sqrt {(c - 1)^2 - 2g}
}{2\sqrt {(c - 1)^2 - 2g - \sqrt {(c - 1)^2 - 2g} } }
\end{equation}
\begin{equation}
\label{eq29} \beta _1^0 = \frac{(2\sqrt {c^2 + l_1 c + l_2 } + 2c +
l_1 )(c - \varphi _0^ - )^{\alpha _1 }}{(2\sqrt {a_1 } \sqrt {c^2 +
l_1 c + l_2 } + b_1 c + l_3 )^{\alpha _1 }}
\end{equation}
\begin{equation}
\label{eq30} \beta _2^0 = \frac{(2\sqrt {c^2 + m_1 c + m_2 } + 2c +
m_1 )(c - \varphi _0^ + )^{\alpha _2 }}{(2\sqrt {a_2 } \sqrt {c^2 +
m_1 c + m_2 } + b_2 c + m_3 )^{\alpha _2 }}
\end{equation}
\begin{equation}
\label{eq31} \beta _1 = \ln \frac{(2\sqrt {a^2 + l_1 a + l_2 } + 2a
+ l_1 )(a - \varphi _0^ - )^{\alpha _1 }}{2\sqrt {a_1 } \sqrt {a^2 +
l_1 a + l_2 } + b_1 a + l_3 )^{\alpha _1 }}
\end{equation}
\begin{equation}
\label{eq32} \beta _2 = \frac{(2\sqrt {b^2 + m_1 b + m_2 } + 2b +
m_1 )(b - \varphi _0^ + )^{\alpha _2 }}{(2\sqrt {a_2 } \sqrt {b^2 +
m_1 b + m_2 } + b_2 b + m_3 )^{\alpha _2 }}
\end{equation}
\begin{equation}
\label{eq33} \xi _0^1 = 2\ln (\beta _1 / \beta _1^0 )
\end{equation}
\begin{equation}
\label{eq34} \xi _0^3 = 2\ln (\beta _2^0 / \beta _2 )
\end{equation}
$a$ and $b$ are two constants satisfying $\varphi _1 (0) = \varphi
_2 (0) = a$, $\varphi _3 (0) = \varphi _4 (0) = b$, and $\varphi _0^
- < a < c < b < \varphi _0^ + $.

\noindent (2) If $g_2 (c) \le g \le g_1 (c)$, then Eq.(\ref{eq8})
has a kink-like wave solution $u = \varphi _1 (\xi )$ of implicit
form (\ref{eq13}) and a antikink-like wave solution $u = \varphi _2
(\xi )$ of implicit form (\ref{eq14}).

\begin{figure}[h]
\centering
\subfloat[]{\includegraphics[height=1.6in,width=2.5in]{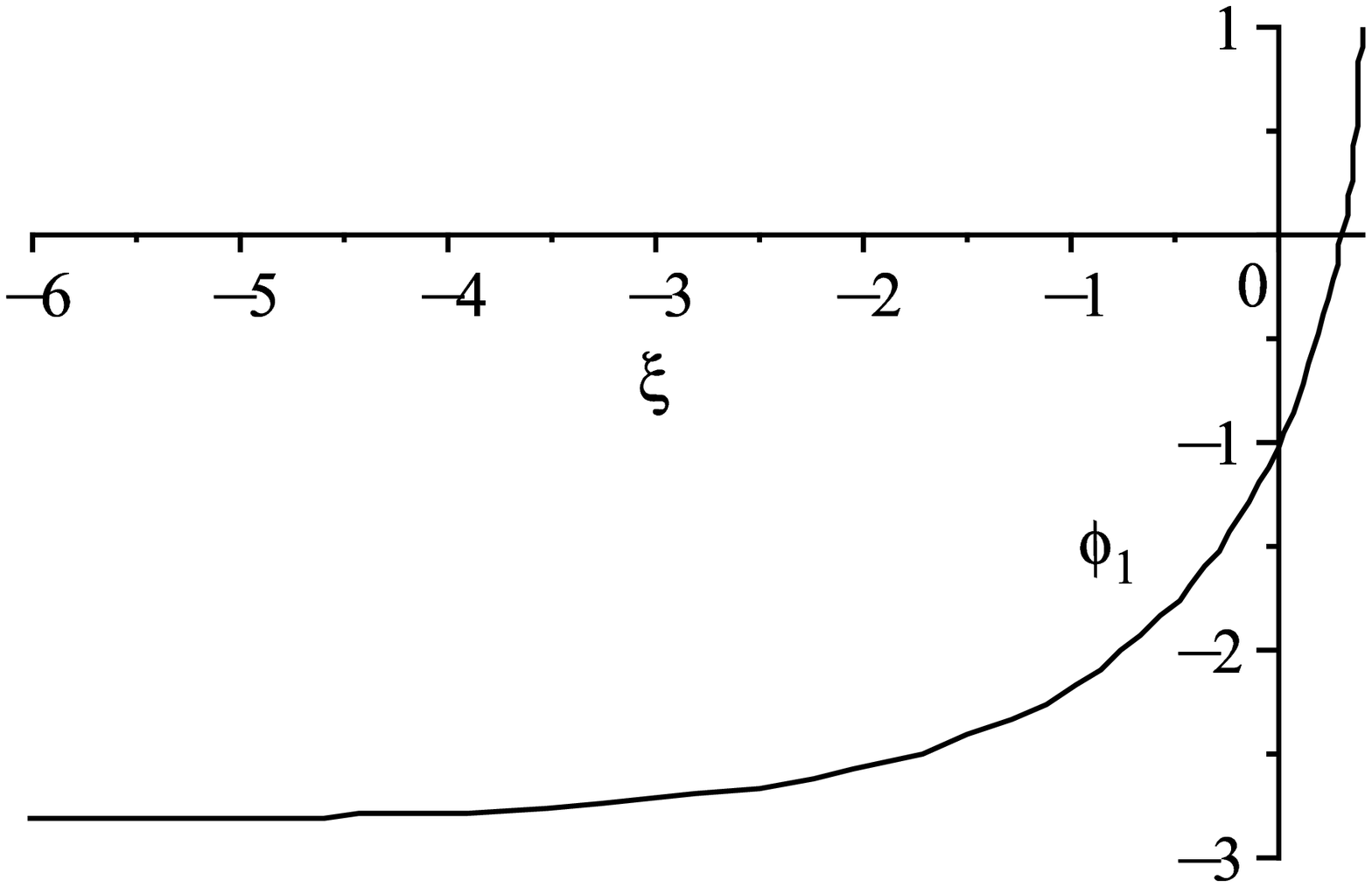}}
\subfloat[]{\includegraphics[height=1.6in,width=2.5in]{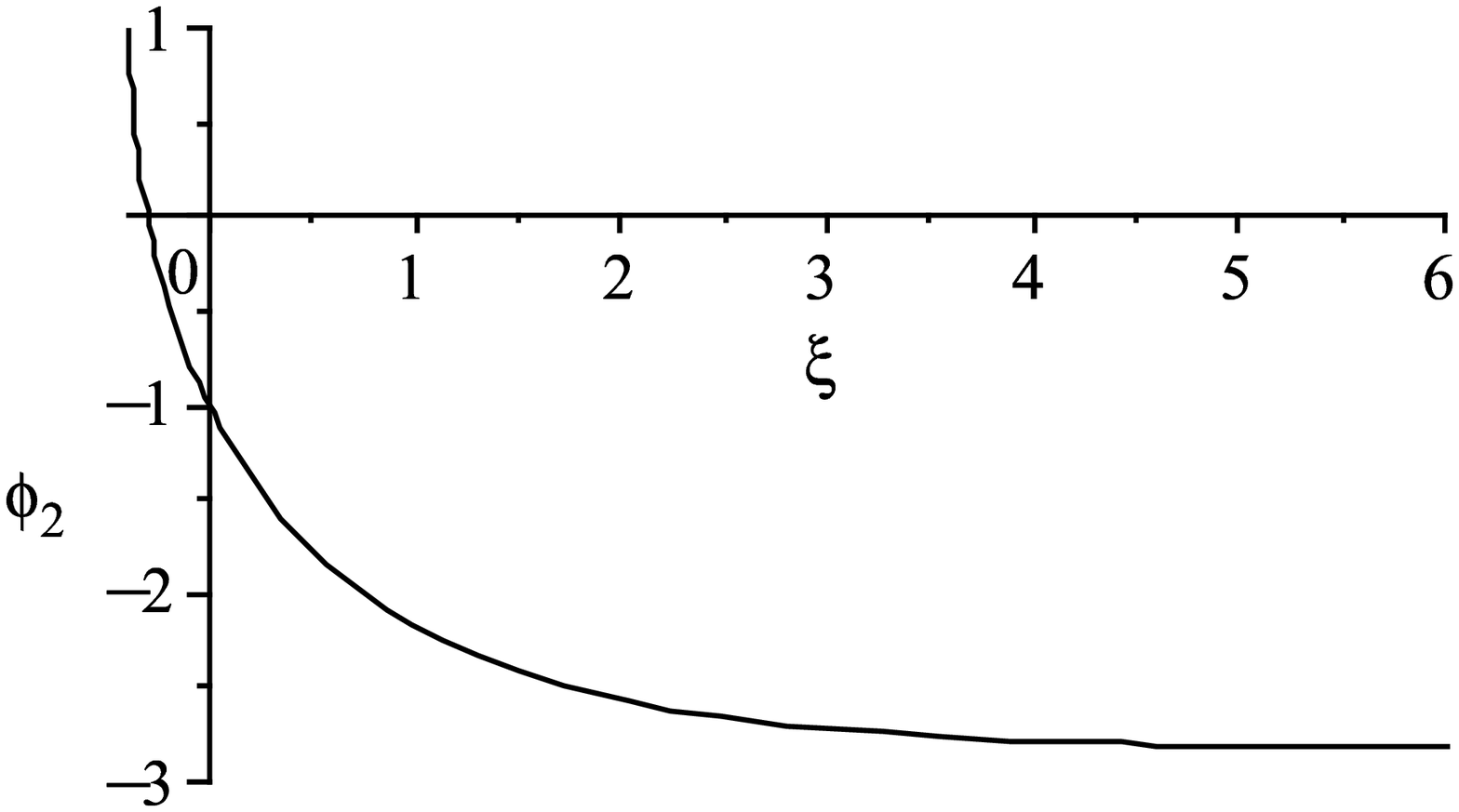}}\\
\subfloat[]{\includegraphics[height=1.6in,width=2.5in]{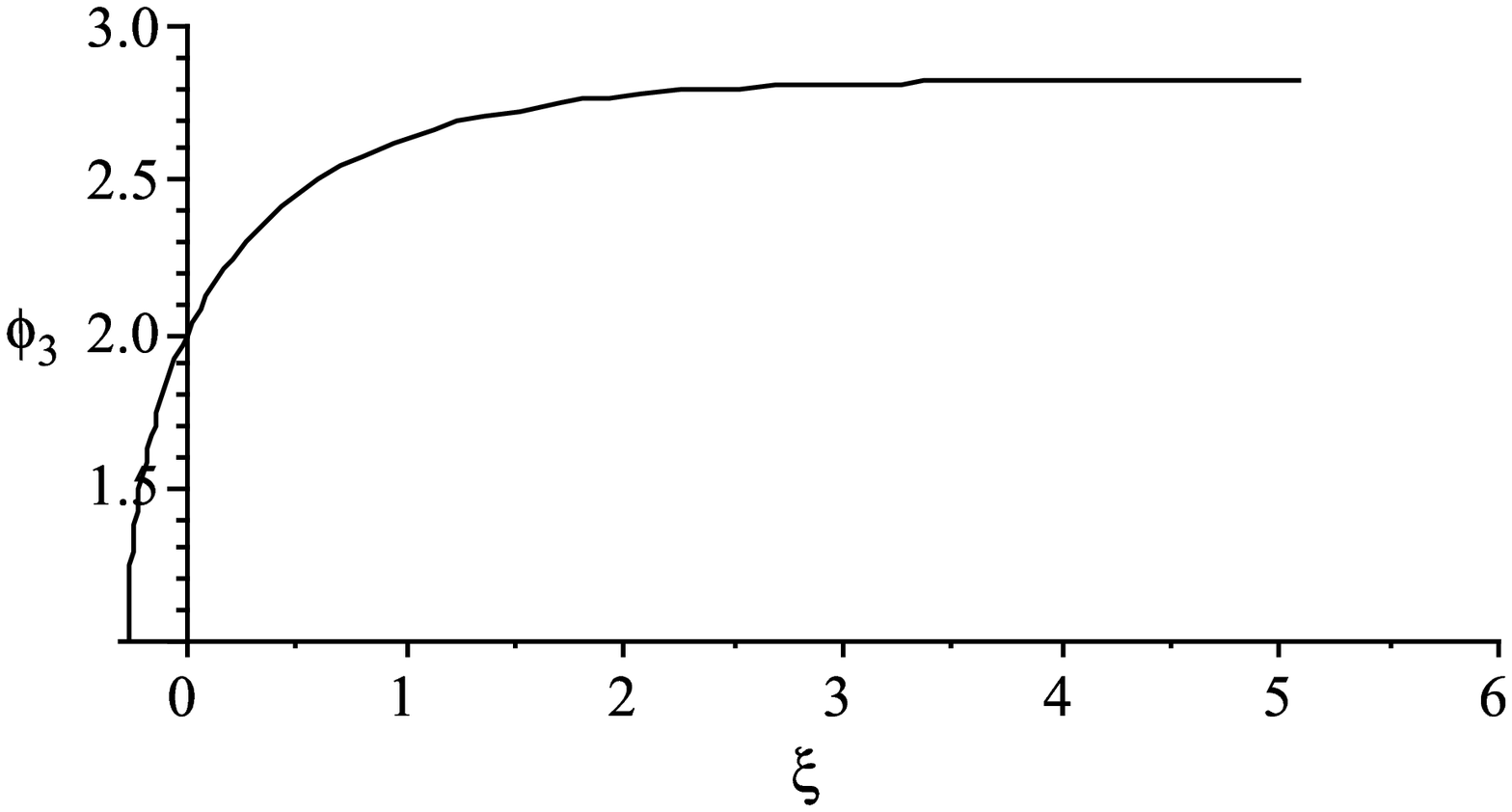}}
\subfloat[]{\includegraphics[height=1.6in,width=2.5in]{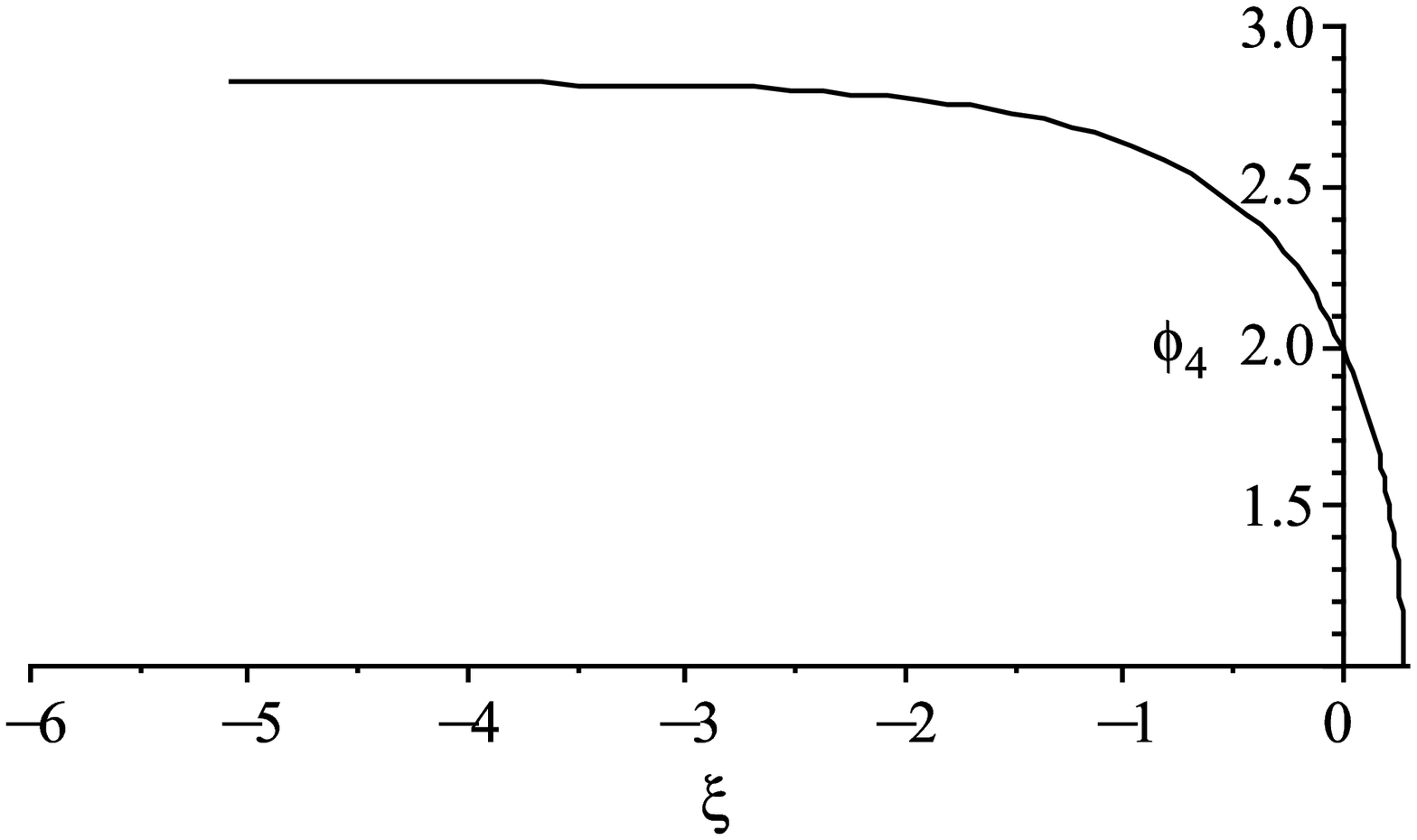}}
\caption{The graphs of $\varphi _i (\xi )(i = 1,\,2,\,3,\,4)$ when
$c = 1$, $g = - 4$, $a = - 1$, $b = 2$.}
\end{figure}
 We will give the proof of this theorem in Section 3.
Now we take a set of data and employ Maple to display the graphs of
$u = \varphi _i (\xi )(i = 1,\,2,\,3,\,4)$.

\noindent \textbf{Example 1.} Taking $c = 1$ and $g =-4<g_2(c)$
(corresponding to (1) of Theorem 1), it follows that $\varphi _0^ -
= - 2.82843$, $\varphi _0^ + = 2.82843$, $l_1 = - 6.99019$ and $l_2
= 15.5425$, $l_3 = 50.8562$, $a_1 = 43.3137$, $b_1 = - 12.647$ ,
$\alpha _1 = - 0.581712$. Further, choosing $a = - 1 \in (\varphi
_0^ - ,c)$, we obtain $\xi _0^1 = \mbox{ 0.387475}$. We present the
graphs of the solutions $\varphi _1 (\xi )$ and $\varphi _2 (\xi )$
in Fig.1 (a) and (b) respectively. Meanwhile, we get $m_1 = 4.32352$
and $m_2 = \mbox{ 0.457528}$, $m_3 = \mbox{ 13.1438}$, $a_2 = \mbox{
20.6863}$, $b_2 = \mbox{ 9.98038}$, $\alpha _2 = \mbox{ 0.40201}$.
Further, choosing $b = 2 \in (c,\varphi _0^ + )$, we get $\xi _0^3 =
\mbox{ 0.274787}$. The graphs of the solutions $\varphi _3 (\xi )$
and $\varphi _4 (\xi )$ are presented in Fig.1(c) and (d)
respentively. The graphs in Fig.1 show that $\varphi _1 (\xi )$ and
$\varphi _3 (\xi )$ are kink-like waves and $\varphi _2 (\xi )$ and
$\varphi _4 (\xi )$ are antikink-like waves.

\section{ Proof of main results}
\setcounter {equation}{0}

Let $u = \varphi (\xi )$ with $\xi = x - ct$ be the solution for
Eq.(\ref{eq8}), then it follows that
\begin{equation}
\label{eq35}
 - c\varphi' + c\varphi ''' + \varphi ' = \varphi \varphi ''' - \varphi
\varphi ' + 3\varphi '\varphi ''
\end{equation}

Integrating (\ref{eq35}) once we have
\begin{equation}
\label{eq36} \varphi ''(\varphi - c) = g - c\varphi + \varphi +
\frac{1}{2}\varphi ^2 - (\varphi ')^2
\end{equation}
\noindent where $g$ is the integral constant.

Let $y = \varphi '$, then we get the following planar system
\begin{equation}
\label{eq37} \left\{ {\begin{array}{l}
 \frac{d\varphi }{d\xi } = y \\
 \frac{dy}{d\xi } = \frac{g - c\varphi + \varphi + \frac{1}{2}\varphi ^2 -
y^2}{\varphi - c}
 \end{array}} \right.
\end{equation}
\noindent with a first integral
\begin{equation}
\label{eq38} H(\varphi ,y) = (\varphi - c)^2[y^2 - \frac{(\varphi -
c)^2}{4} - \frac{2}{3}(\varphi - c) - g - c + \frac{c^2}{2}] = h
\end{equation}
\noindent where $h$ is a constant.

Note that (\ref{eq37}) has a singular line $\varphi = c$. To avoid
the line temporarily we make transformation $d\xi = (\varphi -
c)d\zeta $. Under this transformation, Eq.(\ref{eq37}) becomes
\begin{equation}
\label{eq39} \left\{ {\begin{array}{l}
 \frac{d\varphi }{d\zeta } = (\varphi - c)y \\
 \frac{dy}{d\zeta } = g - c\varphi + \varphi + \frac{1}{2}\varphi ^2 - y^2
\\
 \end{array}} \right.
\end{equation}
The Eq.(\ref{eq37}) and Eq.(\ref{eq39}) have the same first integral
as (\ref{eq38}). Consequently, system (\ref{eq39}) has the same
topological phase portraits as system (\ref{eq37}) except for the
straight line $\varphi = c$. Obviously, $\varphi = c$ is an
invariant straight-line solution for system (\ref{eq39}).

Now we consider the singular points of system (\ref{eq39}) and their
properties . Note that for a fixed $h$, (\ref{eq38}) determines a
set of invariant curves of (\ref{eq39}). As $h$ is varied
(\ref{eq38}) determines different families of orbits of (\ref{eq39})
having different dynamical behaviors. Let $M(\varphi _e ,y_e )$ be
the coefficient matrix of the linearized system of (\ref{eq39}) at
the equilibrium point $(\varphi _e ,y_e )$, then
\[
M(\varphi _e ,y_e ) = \left( {{\begin{array}{*{20}c}
{\quad\quad y_e } \hfill &&& {\varphi _e - c} \hfill \\
 {\varphi _e - (c - 1)} \hfill &&&{- 2y_e} \hfill \\
\end{array} }} \right)
\]
\noindent and at this equilibrium point, we have
\[
J(\varphi _e ,y_e ) = \det M(\varphi _e ,y_e ) = - 2y_e^2 - (\varphi
_e - c)[\varphi _e - (c - 1)],
\]
\[
p(\varphi _e ,y_e ) = trace(M(\varphi _e ,y_e )) = - y_e.
\]
By the theory of planar dynamical system (see \cite {14}), for an
equilibrium point of a planar dynamical system, if $J < 0$, then
this equilibrium point is a saddle point; it is a center point if $J
> 0$ and $p = 0$; if $J = 0$ and the Poinc\'{a}re index of the
equilibrium point is 0, then it is a cusp.

Since system (\ref{eq37}) has the same topological phase portraits
as system (\ref{eq39}) except for the straight line $\varphi = c$.
By investigating the topological dynamics of system (\ref{eq39}), we
can obtain the following properties for system (\ref{eq37}).

\noindent(1) If $g < g_2 (c)$, then system (\ref{eq38}) has two
equilibrium points $(\varphi _0^ - ,0)$ and $(\varphi _0^ + ,0)$.
They are saddle points and there is inequality $\varphi _0^ - < c -
1 < c < \varphi _0^ + $. In this case, there are four orbits
connecting with $(\varphi _0^ - ,0)$, we use $l_{\varphi _0^ - }^1 $
and $l_{\varphi _0^ - }^2 $ to denote the two orbits lying on the
right side of $(\varphi _0^ - ,0)$ (see Fig.2(a)). Meanwhile, there
are four orbits connecting with $(\varphi _0^ + ,0)$, we employ
$l_{\varphi _0^ + }^1 $ and $l_{\varphi _0^ + }^2 $ to denote the
two orbits lying on the left side of $(\varphi _0^ + ,0)$ (see
Fig.2(a)).

\noindent(2) If $g_2 (c) \le g \le g_1 (c)$, then system
(\ref{eq38}) has two equilibrium points $(\varphi _0^ - ,0)$,
$(\varphi _0^ + ,0)$. $(\varphi _0^ - ,0)$ is a saddle point and
$(\varphi _0^ + ,0)$ is a center point or a degenerate center point.
$\varphi _0^{-} $ and $\varphi _0^ + $ satisfy that $\varphi _0^ - <
c - 1 < \varphi _0^ + < c$. $l_{\varphi _0^ - }^1 $ and $l_{\varphi
_0^ - }^2 $ are used to denote the two orbits lying on the right
side of $(\varphi _0^ - ,0)$ (see Fig.2(b))

\noindent(3) If $g_1 (c) < g$, then system (\ref{eq38}) has no
equilibrium point.
\begin{figure}[h]
\centering
\subfloat[$g<g_2(c)$]{\includegraphics[height=1.7in,width=2.5in]{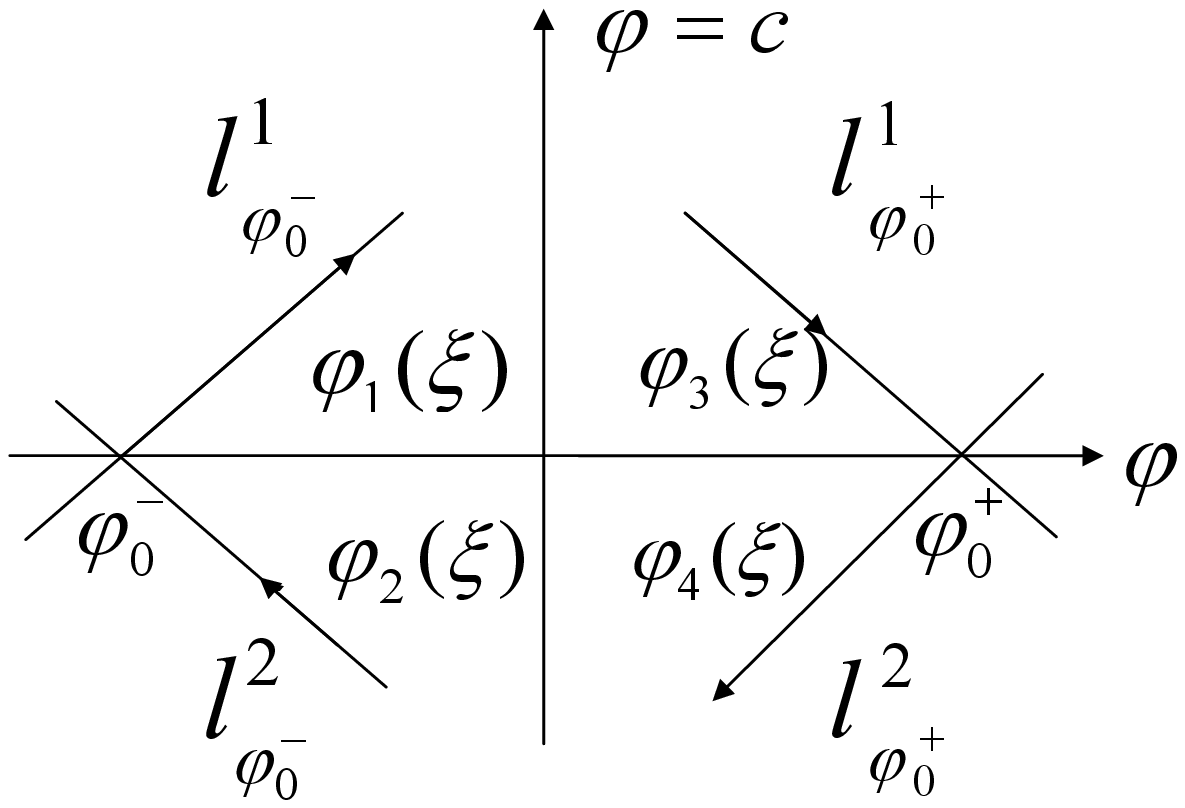}}\hspace{0.1\linewidth}
\subfloat[$g_2 (c) \le g \le g_1
(c)$]{\includegraphics[height=1.7in,width=1.7in]{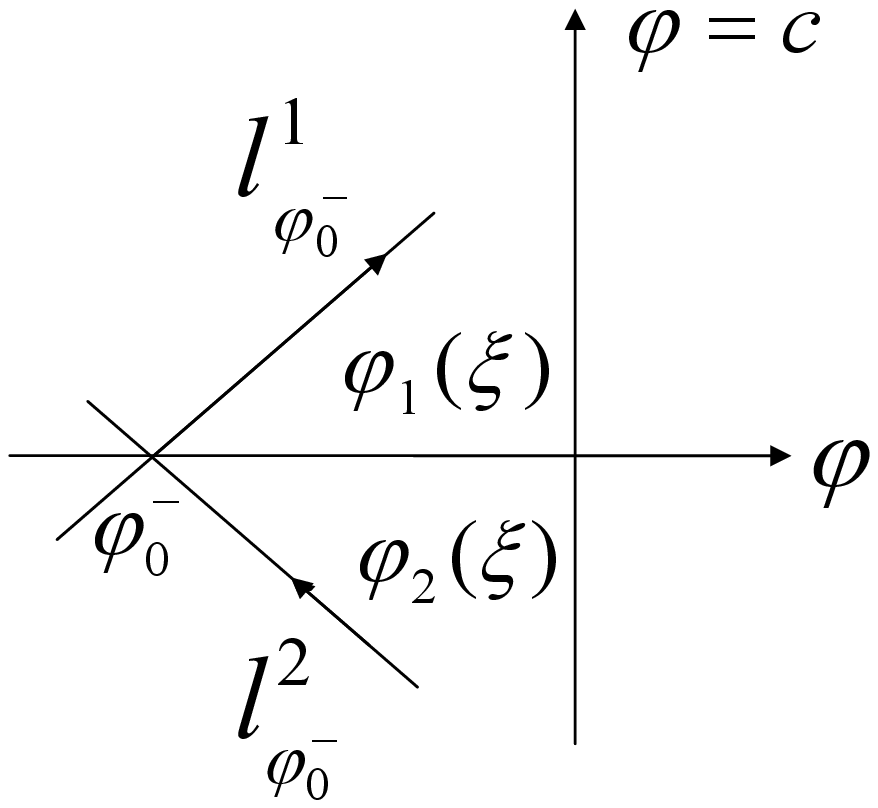}}
\caption{The sketches of orbits connecting with saddle points.}
\end{figure}

On the $\varphi - y$ plane, the orbits $l_{\varphi _0^ - }^1 $,
$l_{\varphi _0^ - }^2 $, $l_{\varphi _0^ + }^1 $ and $l_{\varphi _0^
+ }^2 $ have the following expressions respectively.
\begin{equation}
\label{eq40} l_{\varphi _0^ - }^1 :\quad y =
\frac{1}{2}\frac{(\varphi - \varphi _0^ - )\sqrt {\varphi ^2 + l_1
\varphi + l_2 } }{c - \varphi }
\end{equation}
\begin{equation}
\label{eq41} l_{\varphi _0^ - }^2 :\quad y =
\frac{1}{2}\frac{(\varphi _0^ - - \varphi )\sqrt {\varphi ^2 + l_1
\varphi + l_2 } }{c - \varphi }
\end{equation}
\begin{equation}
\label{eq42} l_{\varphi _0^ + }^1 :\quad y =
\frac{1}{2}\frac{(\varphi _0^ + - \varphi )\sqrt {\varphi ^2 + m_1
\varphi + m_2 } }{\varphi - c}
\end{equation}
\begin{equation}
\label{eq43} l_{\varphi _0^ + }^2 :\quad y =
\frac{1}{2}\frac{(\varphi - \varphi _0^ + )\sqrt {\varphi ^2 + m_1
\varphi + m_2 } }{\varphi - c}
\end{equation}
\noindent where $\varphi _0^ - $ and $\varphi _0^ + $ are in
(\ref{eq10}), $l_1 $ and $l_2 $ are in (\ref{eq17}) and
(\ref{eq18}), $m_1 $ and $m_2 $ are in (\ref{eq20}) and
(\ref{eq21}).

Assume that $\varphi = \varphi _1 (\xi )$, $\varphi = \varphi _2
(\xi )$, $\varphi = \varphi _3 (\xi )$ and $\varphi = \varphi _4
(\xi )$ on $l_{\varphi _0^ - }^1 $, $l_{\varphi _0^ - }^2 $,
$l_{\varphi _0^ + }^1 $ and $l_{\varphi _0^ + }^2 $ respectively and
$\varphi _1 (0) = \varphi _2 (0) = a$, $\varphi _3 (0) = \varphi _4
(0) = b$, where $a$ and $b$ are two constants, and $a \in (\varphi
_0^ - ,c)$, $b \in (c,\varphi _0^ + )$. Substituting
(\ref{eq40})-(\ref{eq43}) into the first equation of (\ref{eq37})
and integrating along the corresponding orbits respectively, we have
\begin{equation}
\label{eq44} {\int_a^{\varphi _1 } {\frac{c - s}{(s - \varphi _0^ -
)\sqrt {s^2 + l_1 s + l_2 } }ds = \frac{1}{2}\int_0^\xi {ds} }}
\quad\quad(\textrm{along}\quad l_{\varphi _0^ - }^1 )
\end{equation}
\begin{equation}
\label{eq45} \int_{\varphi_2}^a {\frac{c - s}{(\varphi _0^ - -
s)\sqrt {s^2 + l_1 s + l_2 } }ds = \frac{1}{2}\int_\xi ^0 {ds} }
 \quad\quad(\textrm{along}\quad l_{\varphi _0^ - }^2 )
\end{equation}
\begin{equation}
\label{eq46} \int_{\varphi _3 }^b {\frac{s - c}{(\varphi _0^ + -
s)\sqrt {s^2 + m_1 s + m_2 } }ds = \frac{1}{2}\int_\xi ^0 {ds} }
\quad(\textrm{along}\quad  l_{\varphi _0^ + }^1 )
\end{equation}
\begin{equation}
\label{eq47} \int_b^{\varphi _4 } {\frac{s - c}{(s - \varphi _0^ +
)\sqrt {s^2 + m_1 s + m_2 } }ds = \frac{1}{2}\int_0^\xi {ds} }
\quad(\textrm{along}\quad l_{\varphi _0^ + }^2 )
\end{equation}
Computing the above four integrals we obtain the implicit
expressions of $\varphi _i (\xi )$ as (\ref{eq13})-(\ref{eq16}).

Meanwhile, suppose that $\varphi _1 (\xi ) \to c$ as $\xi \to \xi
_0^1 $, $\varphi _2 (\xi ) \to c$ as $\xi \to - \xi _0^2 $, $\varphi
_3 (\xi ) \to c$ as $\xi \to - \xi _0^3 $, $\varphi _4 (\xi ) \to c$
as $\xi \to \xi _0^4 $, then it follow from \\(\ref{eq44})
-(\ref{eq47}) that

\begin{equation}
\label{eq48} \xi _0^1 = \xi _0^2 = \int_a^c {\frac{c - s}{(s -
\varphi _0^ - )\sqrt {s^2 + l_1 s + l_2 } }ds}
\quad\quad(\textrm{along}\quad l_{\varphi _0^ - }^1 )
\end{equation}

\begin{equation}
\label{eq49} \xi _0^3 = \xi _0^4 = \int_b^c {\frac{s - c}{(s -
\varphi _0^ + )\sqrt {s^2 + m_1 s + m_2 } }ds}
\quad(\textrm{along}\quad l_{\varphi _0^ + }^2 )
\end{equation}
Computing the above two integrals, we get the expressions of $\xi
_0^1 $ and $\xi _0^3 $ as in (\ref{eq33}) and (\ref{eq34}). The
proof is finished.

\section{Numerical simulations}
\setcounter {equation}{0} In this section, we will simulate the
planar graphs of the kink-like and the antikink-like waves.

From Section 3, we see that in the parameter expressions $\varphi =
\varphi (\xi )$ and $y = y(\xi )$ of the orbits of System
(\ref{eq37}), the graph of $\varphi (\xi )$ and the integral curve
of Eq.(\ref{eq36}) are the same. In other words, the integral curves
of Eq.(\ref{eq36}) are the planar graphs of the traveling waves of
Eq.(\ref{eq8}). Therefore, we can see the planar graphs of the
kink-like and the antikink-like waves through the simulation of the
integral curves of Eq.(\ref{eq36}).

\noindent \textbf{Example 2.} Take the same data as Example 1, that
is $c = 1$, $g = - 4$, $a = - 1$, $b = 2$. Let $\varphi = a = - 1$
in (\ref{eq40}) and (\ref{eq41}), then we can get $y \approx
2.217532085$ or $y \approx - 2.217532085$. And let $\varphi = b = 2$
in (\ref{eq42}) and (\ref{eq43}), then we obtain $y \approx
1.499467912$ or $y \approx-1.499467912$. Thus we take the initial
conditions of Eq.(\ref{eq36}) as follows: (i) Corresponding to
$l_{\varphi _0^ - }^1 $ we take $\varphi (0) = - 1$ and $\varphi
'(0) = 2.217532085$. (ii) Corresponding to $l_{\varphi _0^ - }^2 $
we take $\varphi (0) = - 1$ and $\varphi '(0) = - 2.217532085$.
(iii) Corresponding to $l_{\varphi _0^ + }^1 $ we take $\varphi (0)
= 2$ and $\varphi '(0) = 1.499467912$. (iv) Corresponding to
$l_{\varphi _0^ + }^2 $ we take $\varphi (0) = 2$ and $\varphi '(0)
= - 1.499467912$.
\begin{figure}[h]
\centering \subfloat[$\varphi (0) = - 1$, $\varphi
'(0)=2.217532085$.]
{\includegraphics[height=1.8in,width=2.5in]{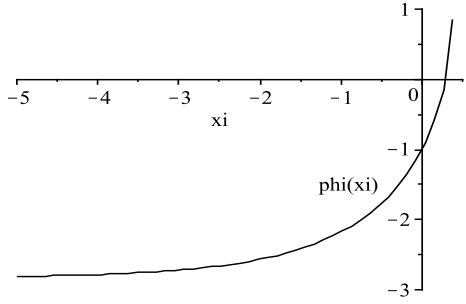}}\hspace{0.06\linewidth}
\subfloat[$\varphi (0) =- 1$, $\varphi '(0) = - 2.217532085$.]
{\includegraphics[height=1.8in,width=2.5in]{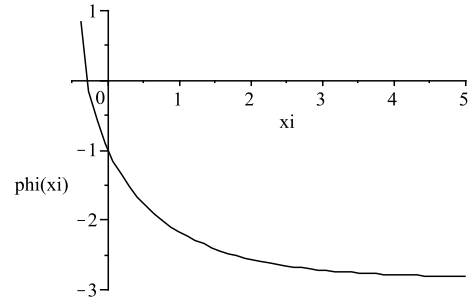}}\\
\subfloat[ $\varphi (0) = 2$, $\varphi '(0) = 1.499467912$.]
{\includegraphics[height=1.5in,width=2.4in]{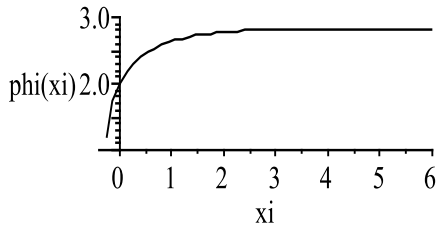}}\hspace{0.07\linewidth}
\subfloat[$\varphi (0) = 2$, $\varphi '(0) = -
1.499467912$.]
{\includegraphics[height=1.45in,width=2.4in]{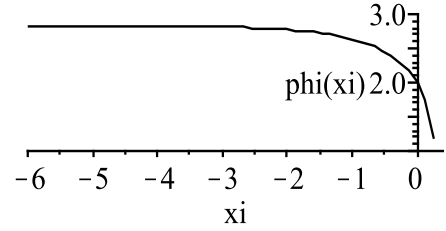}}
\caption{The numerical simulations of integral curves of
Eq.(\ref{eq36}).}
\end{figure}

Under each set of initial conditions we use Maple to simulate the
integrals curve of Eq.(\ref{eq36}) as Fig.3. Comparing Fig.1 with
Fig.3, we can see that the graphs of $\varphi _i(\xi) (i =
1,\;2,\;3,\;4)$  are the same as the simulation of integrals curve
of Eq.(\ref{eq36}). This implies that our theoretic results agree
with the numerical simulations.

\section{Conclusion}

In this paper, we find new bounded waves for the Fornberg-Whitham
equation (\ref{eq8}). Their implicit expressions are obtained in
(\ref{eq13})-(\ref{eq16}). From the graphs (see Fig.1) of the
implicit functions and the numerical simulations (see Fig.3) we see
that these new bounded solutions are defined on some semifinal
bounded domains and possess properties of kink waves and anti-kink
waves.

\end{document}